\documentstyle[preprint,epsfig,aps]{revtex}
\begin{document}

\begin{center}

\begin{large}

{\bf  Probability distribution and sizes of spanning clusters at the 
percolation thresholds} 

\end{large}

\vskip 1cm

 Parongama Sen

Intitute for Theoretical  Physics,
	
University of Cologne

 50923 K\"oln, Germany

\end{center}
\vskip 4cm
\begin{abstract}
For random percolation at $p_c$, the probability  distribution $P(n)$ of the number of spanning clusters
$(n)$ has been studied in large scale simulations. The results are 
compatible with  $P(n) \sim
\exp(-an^2) $ for all dimensions. 
We also study the variation of the average size (mass)  of the spanning clusters when 
there are more than one spanning cluster.
 While the average size 
of the spanning clusters scales as usual with $L^D $ where $D = d- \beta/\nu$ for
any number of clusters, 
it shows a smooth decrease as the number of spanning clusters increases.
\end{abstract}

\pagebreak
\baselineskip = 20 pt

\medskip

More than one spanning cluster (when the spanning is 
considered in one direction) at the percolation threshold  
for dimensions below 6 have been shown to exist in a number of 
recent studies [1-6].    
The numerical evidence  for nonunique  spanning cluster was already there  in 
five dimensions in de Arcangelis' study of spanning clusters [1] nearly 
a decade ago,
 which   was,  however, interpreted as   a finite size effect.

Recently, there have been  some controversies regarding  the behaviour  
of the probability distribution $P(n)$  of the number of spanning 
clusters $(n)$. 
 Using mathematical arguments, Aizenman  has  shown [2,5] that 
$P(n) \sim \exp(-an^2) $ in two dimensions. 
 Aizenman [2] also conjectures that in any dimension
$P(n) \sim \exp(-a n^{d\over{d-1}})$. In simulations reported  earlier  
[3], we had  remarked that the distribution appears to be a simple
 exponential (for 4 and 5 dimensions, only for which considerable data
was available). 
 But for reasons given below and on the basis of our present results, 
this now seems less clear. In the  earlier 
study ([3], referred to as I henceforth), we obtained these probabilities by 
varying the site concentration
 $p$ (the initial value taken close to
the known value of $p_c$) until  a spanning cluster occurs and then 
counted   
 the number $n$ of  such clusters. Here we  use the $p_c$ for the
infinite lattice thereby  increasing the efficiency of the program.
However, the numerical results are then
different. We also later  realised  that the averaging in the 
previous data  in I was actually being taken over a number 
of configurations lesser by a factor of 136 than the 
intended number due to an error in the parallelisation of the program.
However, the main results of I have been verified to be qualitatively 
correct;   the probability of getting more than one
spanning cluster does remain finite. In fact, the present definition  indicates 
a higher  probability with lesser error and as an example we 
have presented the result for two dimensions in fig. 1 for  
this probability. Here, as in I, the probabilities have been
normalised by a factor of $(1-P(0))$, which is  the spanning probability.

Here also, we have simulated hypercubic lattices with   $L^d$  sites in 
$d$ dimensions with 
helical boundary condition.  
The number of configurations over which averaging is done varies from 
$10^4$ to $10^6$ according to the sizes of the lattice. 
Simulations are carried on for 2 to 5 dimensions.

 We have plotted (fig. 2 and 3) the logarithm of the probabilities against 
(a) $n$ and (b) $n^2$ for dimensions 2 to 5 and also against (c) 
 $n^{d\over{d-1}}$ for dimensions 4 and 5.  It is 
observed that   plot (b)  gives a straight line over 
a wider range of $n$  in comparison to
(a) and (c), the latter  correspond to a simple exponential   and
the exponent conjectured in [2] respectively.  
We also   get support for an exponent equal to 2 
 by a different analysis. Assuming that 
$P(n) \sim \exp(-an^\gamma )$,  let 

$$ z = \ln(P(n))/\ln(P(n+1) \sim (n/(n+1))^\gamma \eqno(1)$$

We have first plotted double-logarithmically $ z$ against 
${n\over {n+1}}$ to see
whether we really get a straight line. This has been done for 
five dimensions only, where we have obtained the maximum number of 
spanning clusters. It is difficult to  determine from fig. 4a  
which  straight line  of the three (with slopes = 1.00, 1.25 and 2.00) is the 
best fit to the data (for ${n\over {n+1}} \rightarrow 1$).  
This is  compatible  with the fact that 
there are some regions in  each of the   figures 2a, 2b, 3a, 3b and 3c 
where a straight line can be fitted. However, it is also true that
the data represented by the $\Box $  and $\times $ in fig. 4a,  
have a slope close to $\gamma = 2$,
for which the we have the   best statistics ($L = 15$ and $L =13$).

We have also calculated  $x = \ln(z)/(\ln(n/n+1))$  for different 
sizes which should approach $\gamma$ for $n \rightarrow \infty $ according 
to (1). 
In two dimensions, we do not have enough clusters to 
calculate this quantity.
In order to check whether the values are affected by the 
presence of a prefactor (which has been neglected in (1)), 
we have also calculated $x$  
with the absolute values  $P(n)$ in eqn (1)  replaced by the
ratios  $P(n)/(1-P(0))$.	
We find that  $x$  approaches 2 in dimensions 3 to 5. 
The results are shown in fig. 4b-d, where we show the
values of $x$ for both cases. The $\diamond$ represents 
the case when the prefactor is included. For three
dimensions, in fig. 4b,    the  data close to  $x$ = 6 
correspond to  $n$ = 1 and the one close to 2  to  $n$ = 2. 
 With no prefactor, the data is above 2 for 
$n$ =1 and slightly below 2 for $n$ = 2.
For $d$ = 4 and 5 (fig. 4c    and fig. 4d), the values of $x$ approach 2 
from below as $n$ is increased
when there is no prefactor and from above  when there 
is a prefactor (in general). 
We notice that the values of $x$ from the two different measures 
come closer as $n$ is increased in all  dimensions and it also
approaches 2 clearly in 4 and 5 dimensions where we have data
for  a larger number of clusters. In fig. 4c, data for $n$ = 1,2 and 3 are 
shown while in fig. 4d, data for    $n$ = 2,3,4,5 and 6 are shown. 
Although there are some  fluctuations in the data, apparently  
 the exponent is closer  to 2 than to 1. 
Our conclusion is therefore that for intermediate $n$ 
the probability distribution behaves as 
a stretched exponential $P(n) = \exp(-an^\gamma )$, with the  exponent close to
 2 in all dimensions.  Apparently, the value of $a$ decreases with 
dimensionality
($\sim 0.5$ for $d = 4 $  
and $\sim 0.2$ for $d = 5 $).   

We have also studied the sizes or masses (i.e., the number of occupied 
sites in the cluster) of the spanning clusters.
The average size of the spanning clusters indicates that even 
when there are more than one cluster, the size still scales as 
$L^D, $ with $ D = d -\beta/\nu $, as in the case of unique cluster [7]. 
However, there is a decrease 
in the average size as more spanning clusters appear and this is 
also expected. In two or three  dimensions, 
where we have at the most 2 or 3 clusters respectively in general, the average 
size of spanning clusters versus the sytem sizes $L$ 
data is shown in fig. 5a. For 4 and 5 dimensions, 
where more clusters are
obtained, the scaled data $(\bar m/L^D)$, where $\bar m$ is the
average mass,  against the cluster number is shown in 
a log-log plot (fig.5b). It is difficult to verify whether the variation with 
$n$ is a power law because of the fluctuations  for large number of
clusters. These fluctuations are unavoidable as the
averaging for cluster sizes for larger $n$ is done over a smaller 
number of cases (as the probabilty decreases with number of 
clusters).  However, the
smooth decrease is clearly 
indicated.

We also calculated the second and third moments of the 
masses where the $q$th moment is defined as 
$m_q$ = ${1\over{n}}\sum_{i}^n(m-\bar m)^q$,
 to get an idea of the probability distribution of the sizes. 
While $m_2/{\bar m^2}$ remains  more or less a constant for 
different system sizes in five dimensions for
six clusters, $m_3/{\bar m^3}$ remains positive and 
fairly constant indicating an asymmetric distribution : possibly
one large cluster and several smaller clusters exist.

\vskip 1cm

 This work is supported by  SFB 341.  The author expresses sincere
gratitude to Dietrich Stauffer for suggestions, discussions and encouragement.
She also thanks A. Aharony and J. P. Hovi for discussions.
 Acknowledgements are also due to
 German Israeli foundation and to 
HLRZ Julich for  $10^4$  hrs on Paragon.

\pagebreak

\noindent {\bf References}
\begin{enumerate}

\item {  } L. de Arcangelis,   J.  Phys. A {\bf 20}, 3057  (1987).

\item {  } M. Aizenman, Nucl. Phys. (in press). 
\item {  }  P. Sen,  Int.  J. Mod. Phys. C {\bf 7}, 603 (1996). 
\item {  } C.-K. Hu and C.-Y. Lin, Phys. Rev. Lett. {\bf 77}, 8 (1996).
\item {  } M. Aizenman,  {\it The IMA Volumes in Mathematics and 
its Applications} (Springer-Verlag, 1996).  
\item {  } A. Klassmann, Masters thesis Cologne University, has 
shown that the mathematical proofs of uniqueness (as cited in [2]) work
best for $P_\infty > 0$, i.e., above $p_c$.
\item {  } D. Stauffer and A. Aharony, {\it Introduction to Percolation
Theory} (Taylor and Francis, London, 1994).

\end{enumerate}
\pagebreak

\noindent {\bf Figure Captions}

\vskip 1cm

\noindent Fig. 1 The variation of the probability of getting more than one
spanning cluster is shown against different system sizes in  two  
 dimensions.

\medskip

\noindent Fig. 2   Probability distribution ($P(n)$) of the number of
spanning clusters ($n$) for  2 ($\diamond ~ L = 150$) and 3 (upper data points) 
dimensions against (a) $n$ and
(b) $n^2$.  The different symbols are for sizes  : 
$+~~ L = 100, ~~\Box ~L~ = 125,
~~\times ~~L = 150, ~~\triangle~~ L = 175,   \ast~~ L = 200$ in 3 dimensions. 
\medskip

\noindent Fig. 3   Probability distribution ($P(n)$) of the number of
spanning clusters ($n$) for  4 (lower data points) and 5 (upper data points) 
  dimensions against (a) $n$ and
(b) $n^2$ and (c) against $n^{d\over{d-1}}$. 
The different symbols correspond to
different system sizes : $\diamond $ for $L$ = 21, 
$+$ for $L$ = 19, $\Box$ for $L$ = 17, 
$\times $ for $L$ = 15 and $\triangle $ for $L$ = 13 for
$d$ = 5 and $\ast $ for $L$ = 39, filled $\Box$ for $L$ = 35, 
filled $\triangle $ for $L$ = 30, $\diamond $ for $L$ = 27, and 
$+$ for
$L$ = 23 for $d=4$. 

\medskip

\noindent Fig.  4a   The values of $z$ against ${n\over{n+1}}$
  as found for 
5 dimensions. The different symbols represent different sizes : ~~$\diamond L =
21, 
~+~~ L = 19, ~~\Box~~ L = 15
$ and $ ~~\times~~ L = 13$. The three straighlines have slope 
= 1.00, 1.25 and 2.00 (from left to right). 
\medskip

\noindent Fig.  4b   The values of $x$ vs. $L$ as found for 
3 dimensions. The $\diamond$ are for the ratios and the $+$ for
the absolute values of $P(n)$. $n$ increases as the value of 
$x$ decreases.
\medskip

\noindent Fig.    4c The values of $x$  vs. $L$ as found for 
4 dimensions. The $\diamond$ are for the ratios and the $+$ for
the absolute values of $P(n)$. $n$ increases as the value of 
$x$ decreases for $\diamond$ and vice versa for the $+$.
\medskip

\noindent Fig.    4d  The values of $x$ vs. $n$ as found for 
5 dimensions. The $\diamond$ are for the ratios and the $+$ for
the absolute values of $P(n)$. 
The data correspond to different sizes $L = 13 $ to $ L = 21 $.
\medskip

\noindent Fig 5a  The variation of the average mass of the
spanning clusters against the system sizes for $d$ = 2 ($\diamond$ for no. of
spanning clusters $n = 1$ and $+$ for $n = 2$) and  $d = 3$  
($\Box $ for $n = 1$, $\times $ for $n = 2$ and $\triangle $ for $n$ = 3).
 The system sizes in $d$ = 3 have been multiplied by
10 in order to be shown in the same range.

\medskip

\noindent Fig 5b  The variation of the scaled average mass $(\bar m/L^D)$ of the
spanning clusters against the the number of clusters
in  4 and 5 dimensions.
 The  average masses in four dimensions have been divided by
2 for better viewing.
The different symbols correspond to
different system sizes : $\diamond $ for $L$ = 21, 
$+$ for $L$ = 19, $\Box$ for $L$ = 17, 
$\times $ for $L$ = 15 and $\triangle $ for $L$ = 13 for
$d$ = 5 and $\ast $ for $L$ = 39, filled $\Box$ for $L$ = 35, 
filled $\triangle $ for $L$ = 30, $\diamond $ for $L$ = 27, and 
$+$ for
$L$ = 23 for $d =4$. 
\psfig {file = 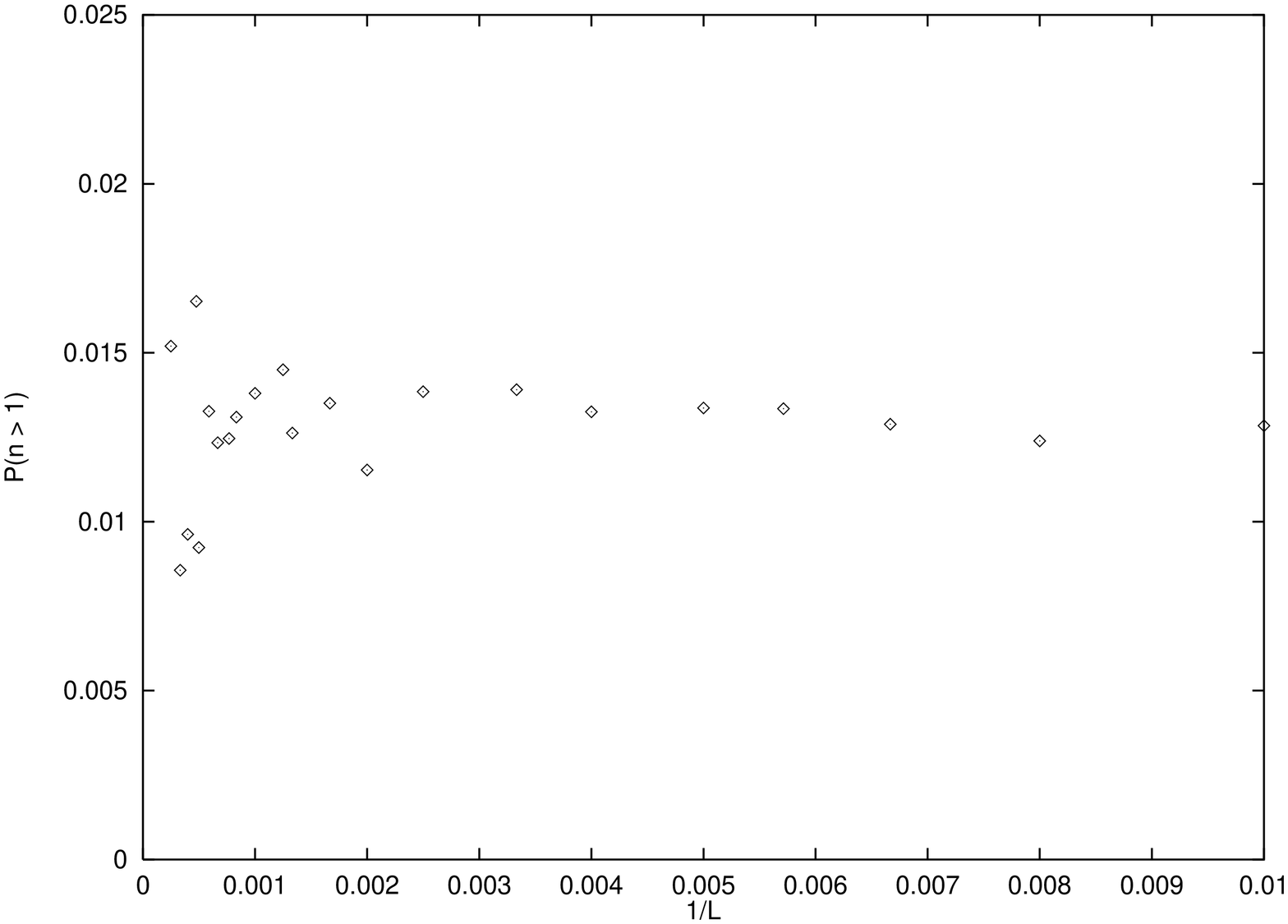, width = 3in, angle = 270}
\psfig {file = 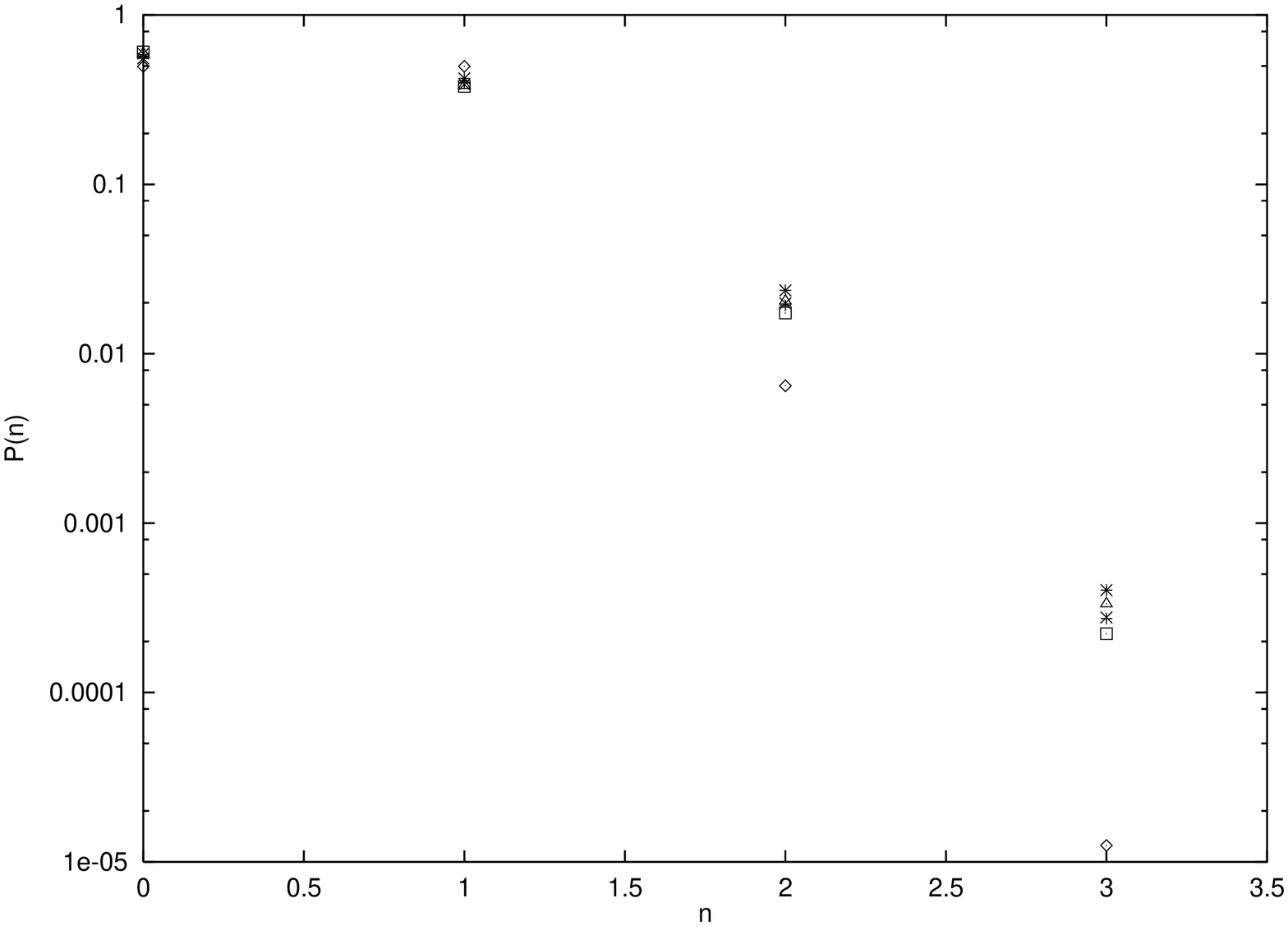, width = 3in, angle = 270}
\psfig {file = 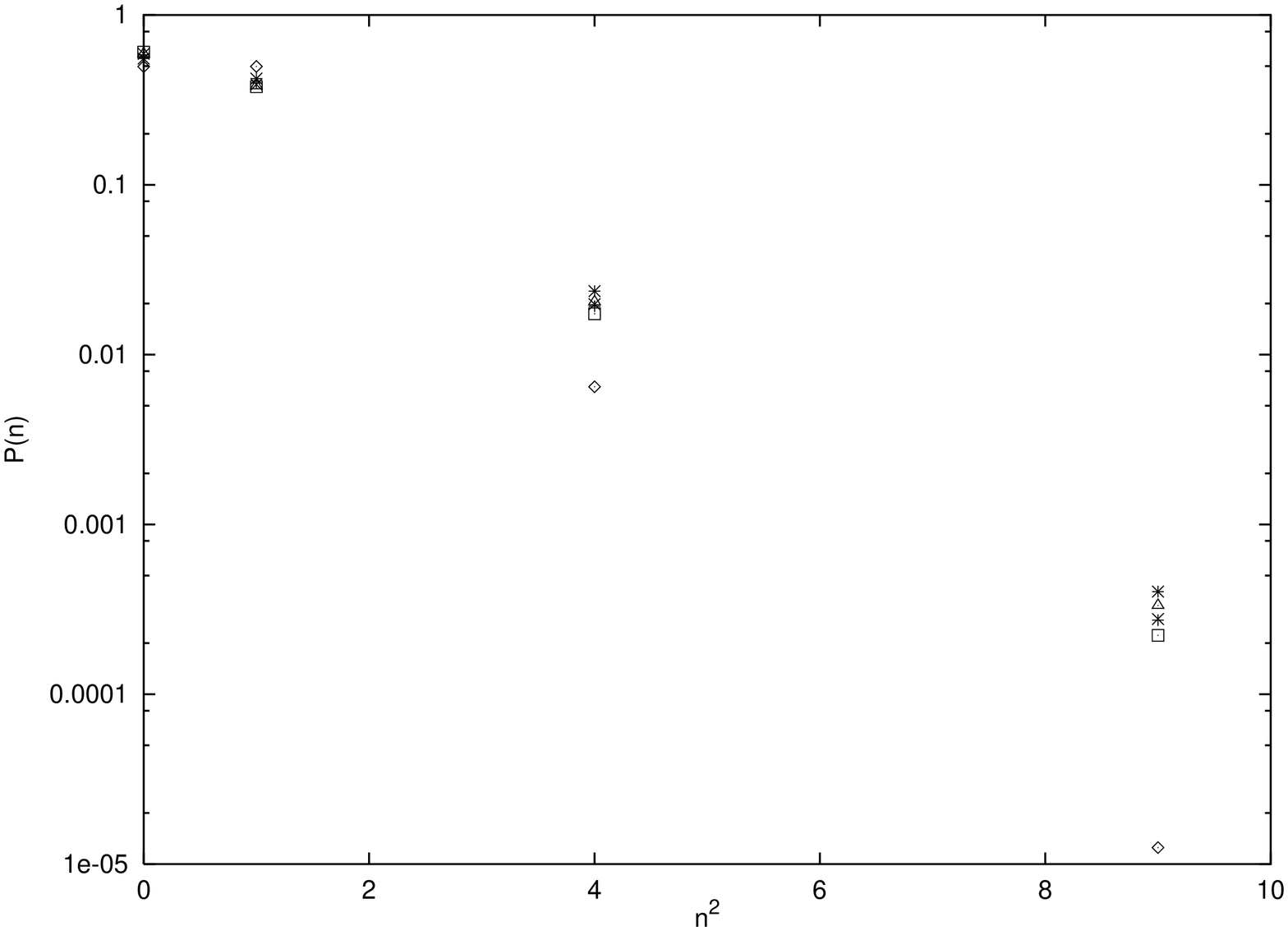, width = 3in, angle = 270}
\psfig {file = 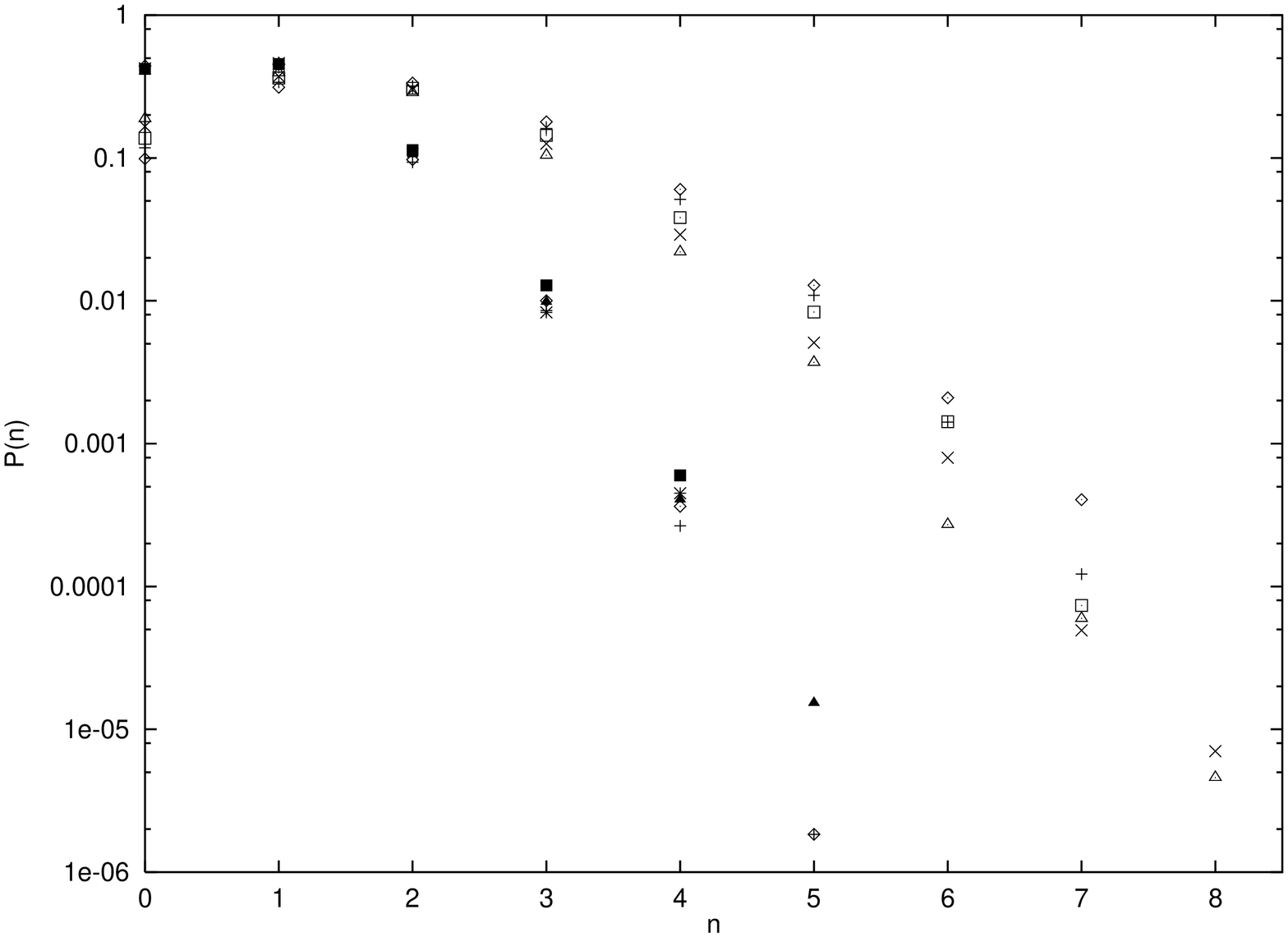, width = 3in, angle = 270}
\psfig {file = 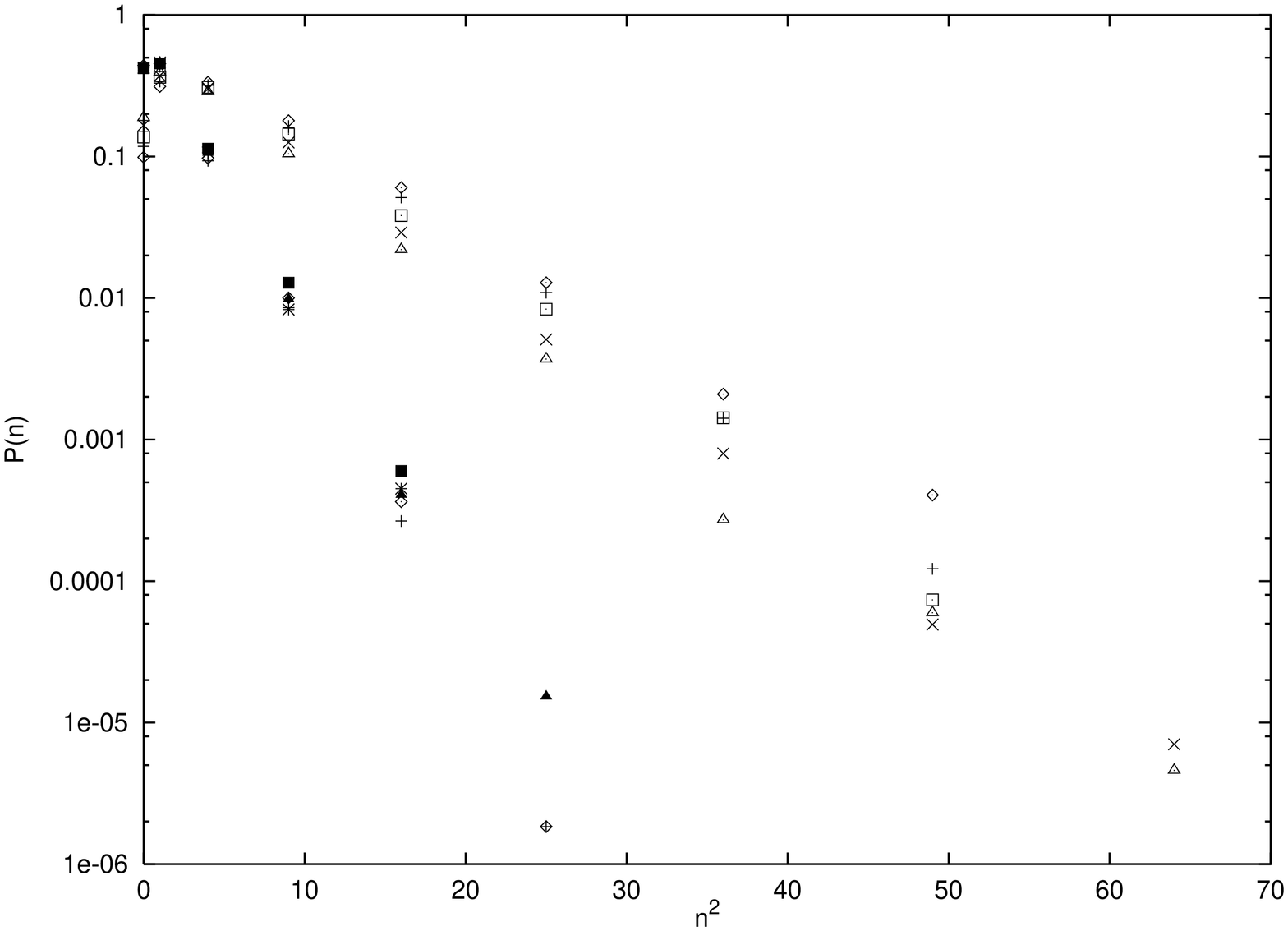, width = 3in, angle = 270}
\psfig {file = 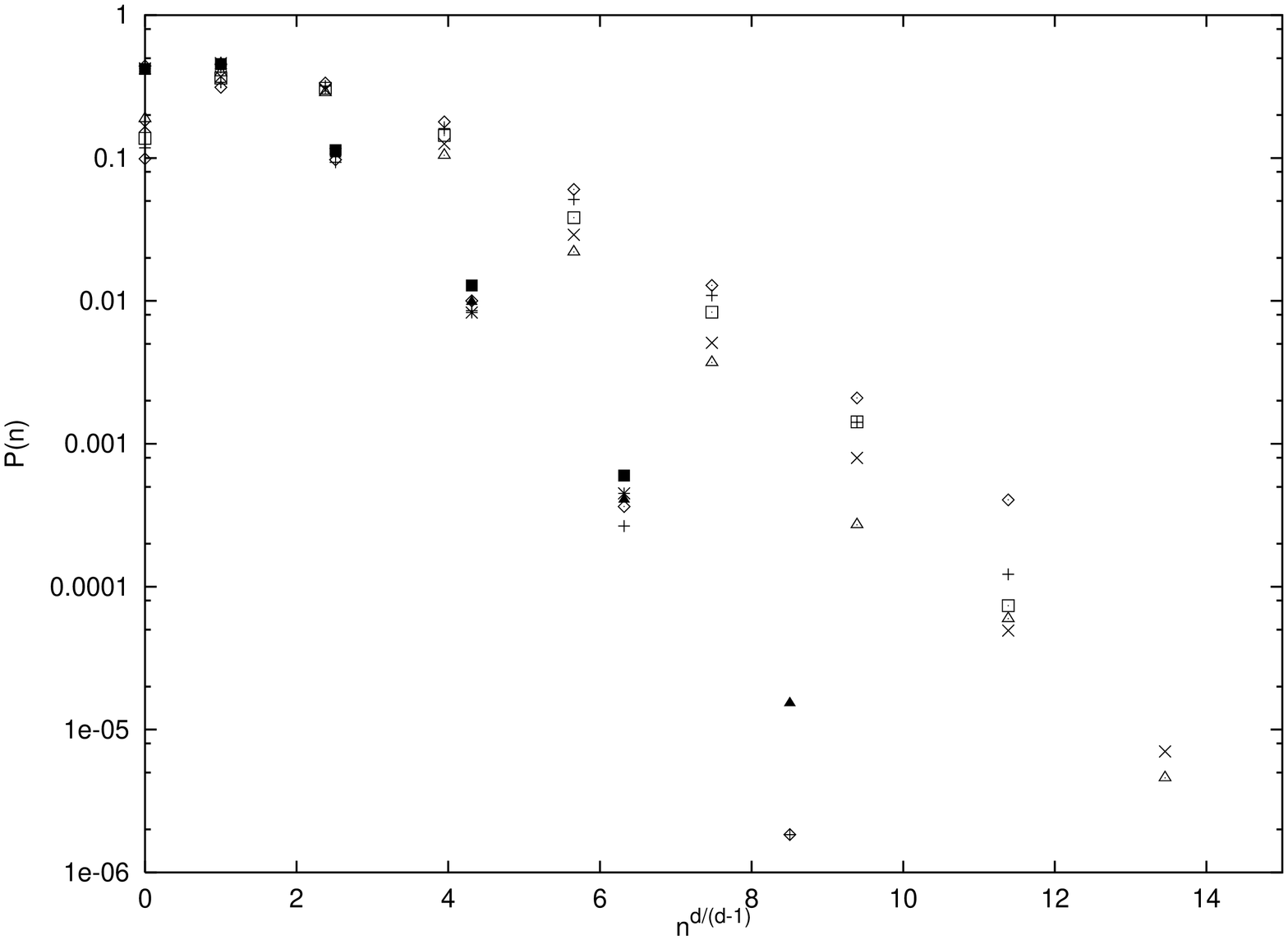, width = 3in, angle = 270}
\psfig {file = 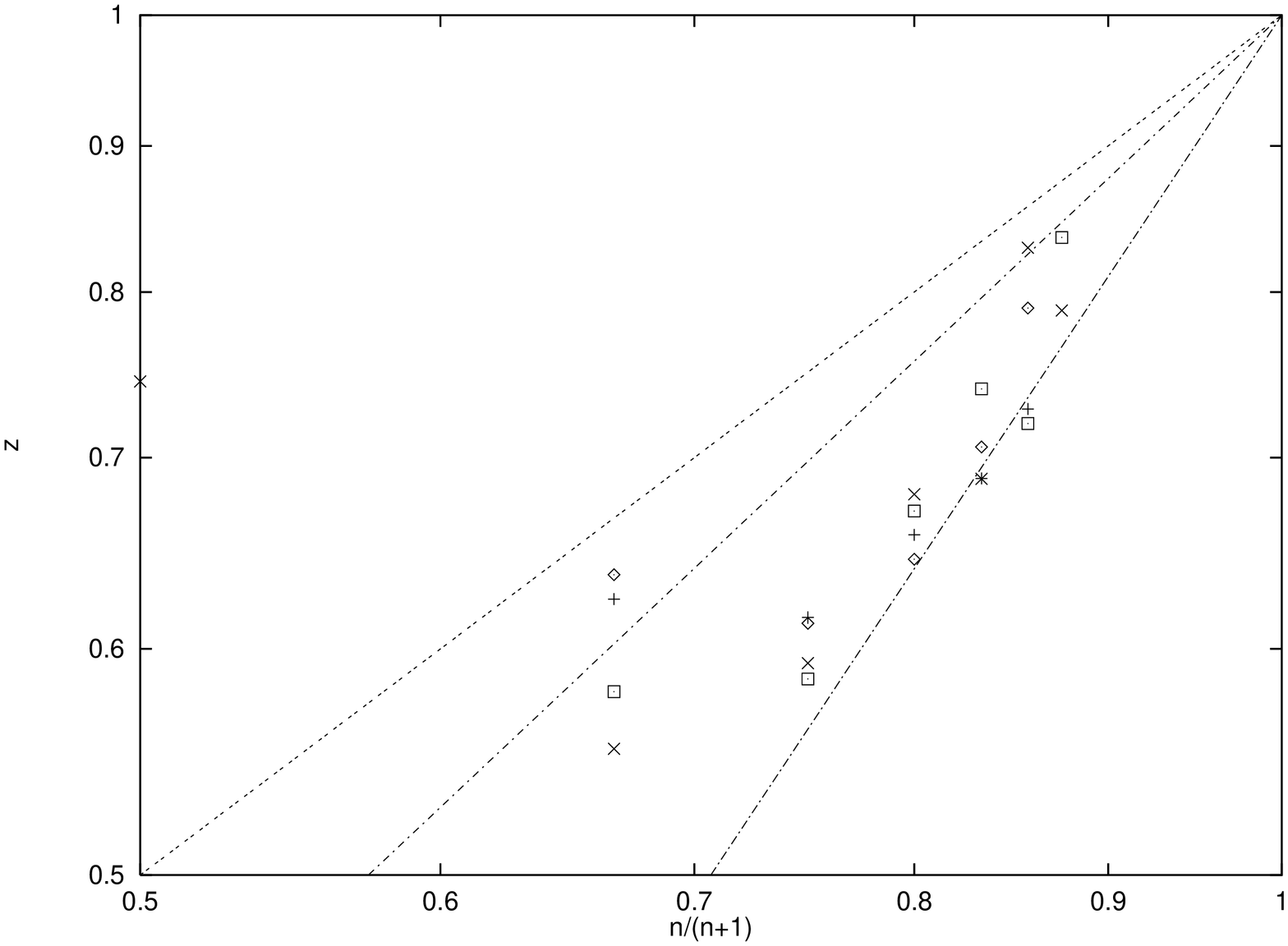, width = 3in, angle = 270}
\psfig {file = 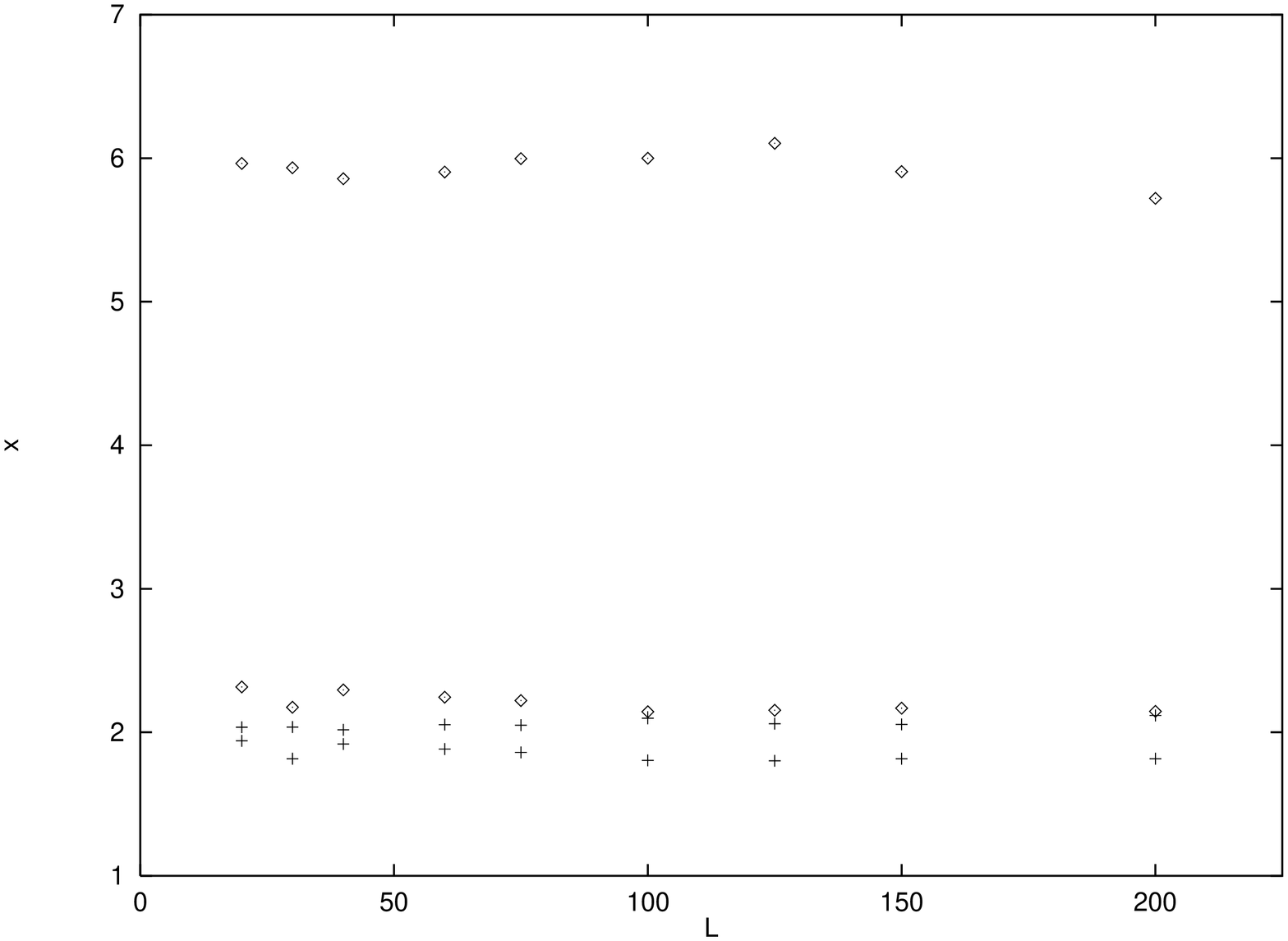, width = 3in, angle = 270}
\psfig {file = 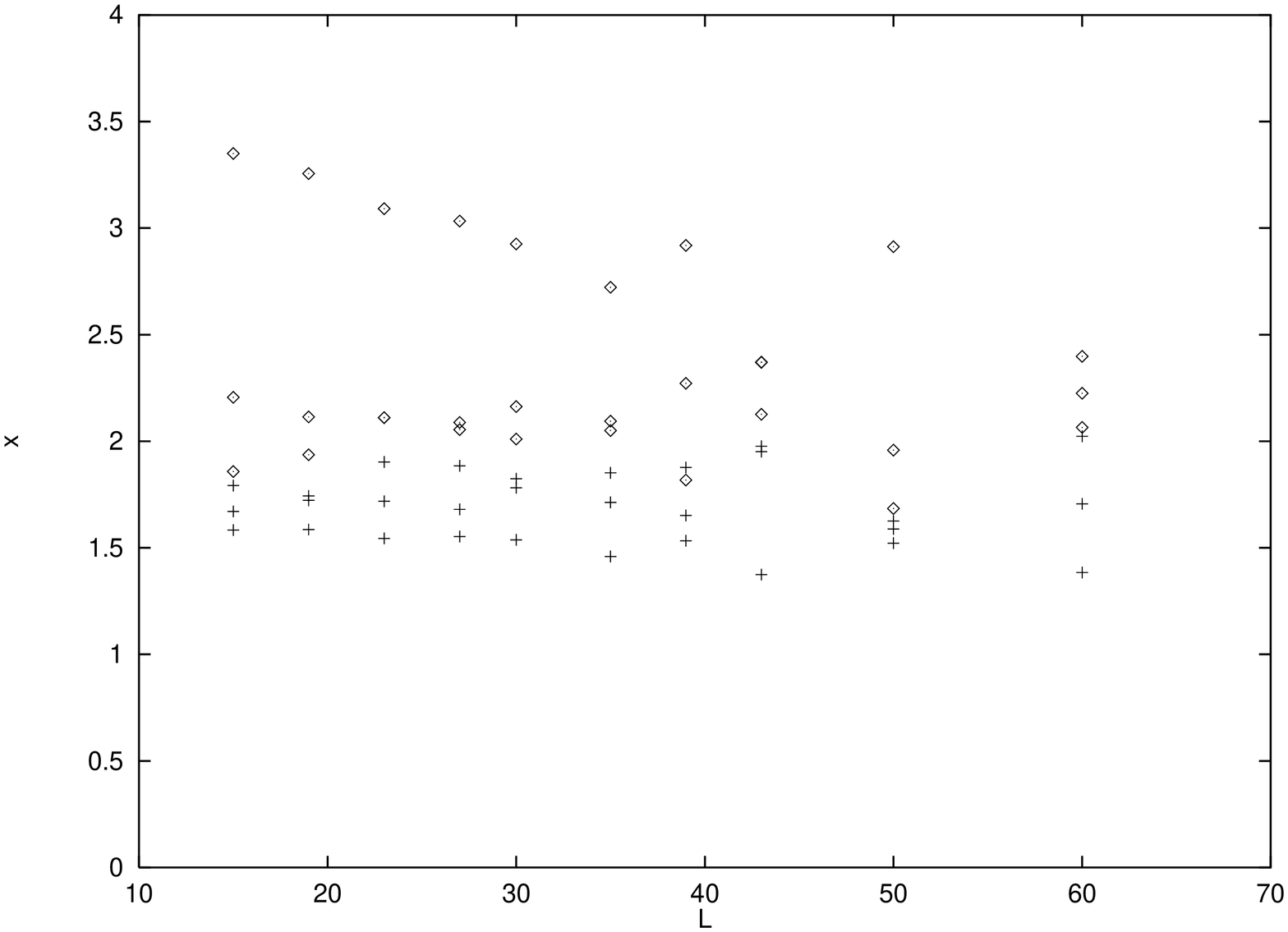, width = 3in, angle = 270}
\psfig {file = 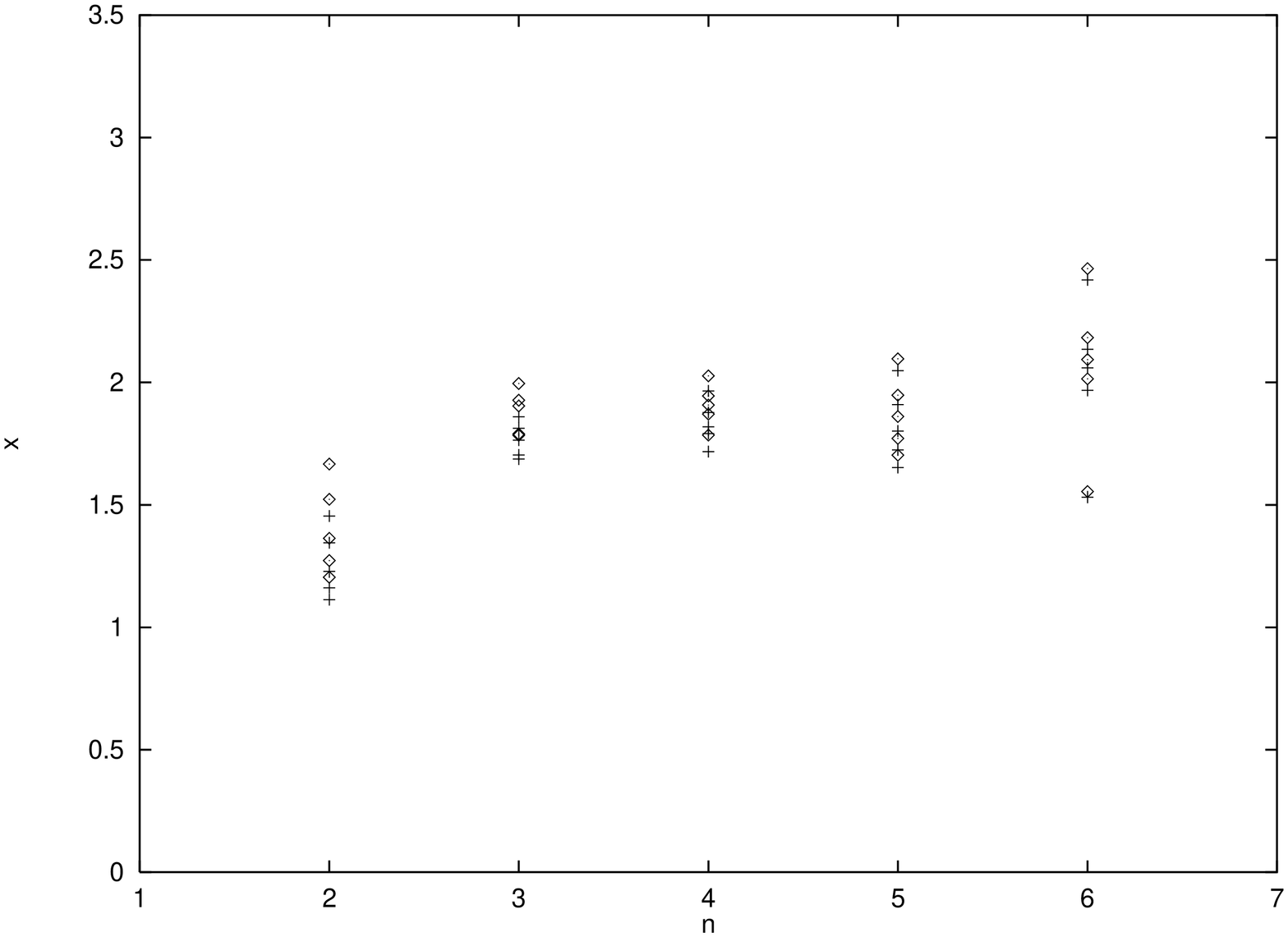, width = 3in, angle = 270}
\psfig {file = 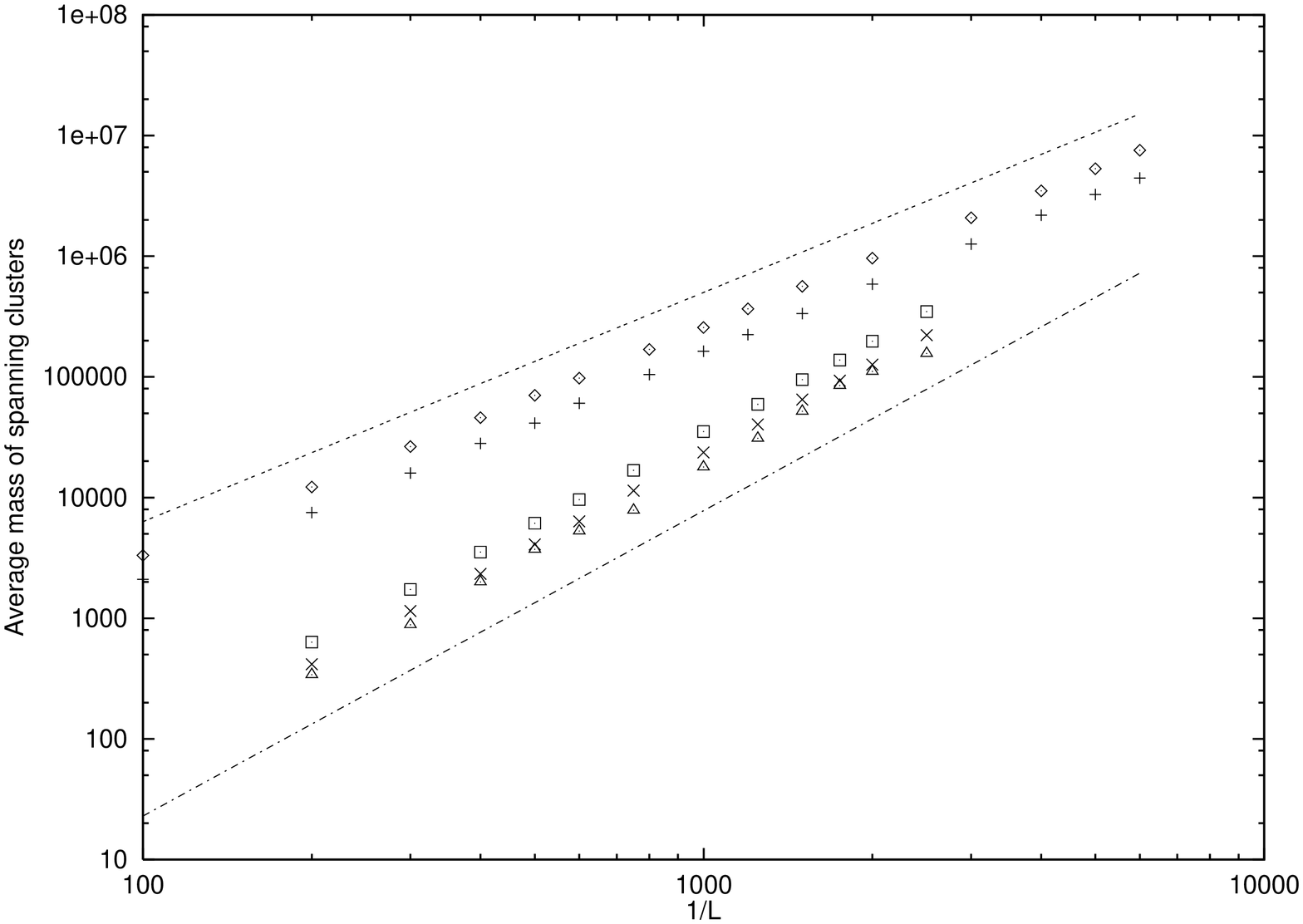, width = 3in, angle = 270}
\psfig {file = 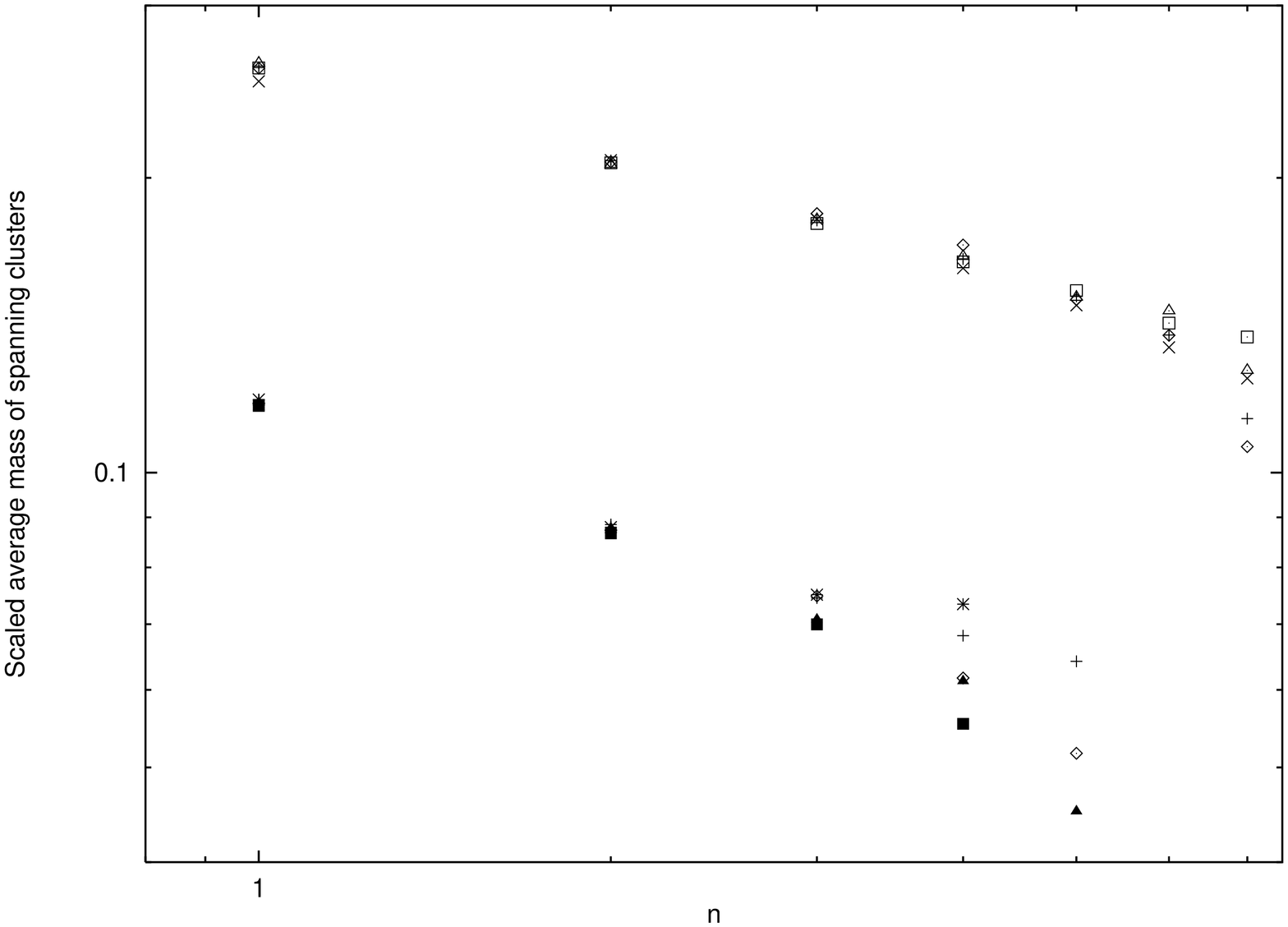, width = 3in, angle = 270}
\end{document}